\documentclass[10pt,twoside]{article}

\def\sech{\mathop{\rm sech}\nolimits}
\def\nn{\nonumber}
\usepackage{graphicx}
\usepackage{epsfig}

\makeatletter
\evensidemargin \oddsidemargin
\makeatother

\begin{document}
\thispagestyle{empty}
\setcounter{page}{1}

\noindent
{\footnotesize {\rm To appear in\\[-1.00mm]
{\em Dynamics of Continuous, Discrete and Impulsive Systems}}\\[-1.00mm]
http:monotone.uwaterloo.ca/$\sim$journal} $~$ \\ [.3in]

%%USE THE STUFF BELOW AS A GUIDE TO SET UP THE START OF PAPER%%

\begin{center}
{\large\bf \uppercase{Solitary waves and their stability in colloidal media: semi-analytical solutions}}

\vskip.20in

T.R. Marchant$^{1}$\ and\ N.F. Smyth$^{2}$ \\[2mm]
{\footnotesize
$^{1}$School of Mathematics and Applied Statistics,\\
The University of Wollongong, Wollongong, NSW 2522, Australia. \\[5pt]
$^{2}$
School of Mathematics and the Maxwell \\
Institute for Mathematical Sciences, \\
        University of Edinburgh, \\ The King's
        Buildings, Mayfield Road, Edinburgh, Scotland, U.K., EH9 3JZ.\\
Corresponding author email: N.Smyth@ed.ac.uk}
\end{center}
{\footnotesize
\noindent

{\bf Dedication}: This paper is dedicated to Professor James Hill on the occasion of his 65th birthday.
Prof.\ Hill has been a source of inspiration to the authors, over many years, due to his significant
contributions in many diverse and important areas within Applied Mathematics.
He has also played a key role in the support and mentoring of a new generation of researchers.
We wish him well in his future endeavours.

\vskip 5mm
\begin{abstract}
Spatial solitary waves in colloidal suspensions of spherical dielectric nanoparticles are considered.
The interaction of the nanoparticles is modelled as a hard-sphere gas, with the
Carnahan-Starling formula used for the gas compressibility.  Semi-analytical solutions, for both one and
two spatial dimensions, are derived using an averaged Lagrangian
and suitable trial functions for the solitary waves. Power versus propagation constant curves and
neutral stability curves are obtained for both cases, which illustrate that multiple solution branches occur
for both the one and two dimensional geometries.  For the one-dimensional case it is found that three solution
branches (with a bistable regime) occur, while for the two-dimensional case two solution branches (with a single
stable branch) occur in the limit of low background packing fractions.  For high background packing fractions the power
versus propagation constant curves are monotonic and the solitary waves stable for all parameter values.  Comparisons
are made between the semi-analytical and numerical solutions, with excellent comparison obtained.
\end{abstract}

{\bf Keywords.}  colloid, solitary wave, modulation theory, stability, bifurcation. \\[3pt]
{\small\bf AMS (MOS) subject classification:} 35, 78.}

\vskip.2in

\section{Introduction}

Over the last two decades the mechanical interaction between light and soft matter has received considerable
attention, including the emergence of new tools in optics, such as optical tweezers and traps, see
\cite{daoud,grier,ashkin}.  In the colloidal medium considered here, which is composed of a suspension of
dielectric nanoparticles, the exceptionally high optical nonlinearity is due to the optical gradient force
changing the concentration or orientation of the colloidal particles. This leads to an intensity-dependent refractive
index change and hence a mutual interaction between the colloidal particles and light.  Possible new applications in
colloidal media include optical sensors or selective particle trapping and manipulation.
Spatial solitary waves form in a colloidal medium due to a balance between diffraction of the light beam and
the nonlinear particle-light interaction.

\cite{optexpr,yuri,poland} derived colloidal media governing equations by assuming that the colloidal suspension
represents a hard-sphere gas.  They used the Carnahan-Starling formula for the compressibility of the hard-sphere
colloid. They then considered one and two dimensional colloidal equations and derived numerically exact propagation
constant versus power curves.  They showed that bistable behaviour occurs for some parameter values and considered
solitary wave interactions for solitary waves of the same power, from the same and different solution branches. They
found dramatically different interaction behaviour for solitons from the same and different branches.  In the 2-D
case, only two solution branches can occur and the bistable behaviour of the 1-D colloidal solitary wave is absent.

\cite{matus} considered a colloidal suspension of two different species of nanoparticles, one with refractive index
higher than the background medium and the other with refractive index lower than the background medium.  One species
is described by a hard-sphere gas with the Carnahan-Starling compressibility formula and the other by an ideal gas.
Numerical solitary wave solutions were found for the governing equations which show that bistability can occur for
the two-dimensional geometry, which is not possible for suspensions composed of only a single nanoparticle species.

\cite{elgan} modelled a colloidal suspension of nanoparticles as an ideal gas and considered two cases, one for which
the refractive index of the nanoparticles was greater than the background medium, and the other for which the refractive
index is lower. The governing equation in the two cases of positive and negative polarizability was an NLS-type equation
with exponential nonlinearity and a form of saturable exponential nonlinearity, respectively.  The stability of the
solitary waves solutions was considered numerically in one and two spatial dimensions.  \cite{elgan09} considered a
power series for the compressibility of the colloid particles with the coefficients of the series found by using
the Debye-Huckel model for the particle interactions.  This interaction model included screening effects due to
the ions in the electrolyte solution.  Solitary wave stability was considered, with one-dimensional waves found to be
stable and two-dimensional waves unstable.  \cite{lee} experimentally considered the nonlinear optical response of a
nanoparticle suspension and found that a higher-order NLS equation with coefficients fitted from a two term series
expansion for the compressibility best matched the experimental data.

The governing equation for wave propagation in colloidal media is an NLS-type equation for which the form of the
nonlinearity in the NLS-type equation depends on the assumed nature of the nanoparticle interactions.  No exact
solitary wave solutions exist for these NLS-type equations, so most existing work
on has been numerical or based on a mix of various asymptotic, approximate and numerical methods.
An effective technique for deriving semi-analytical solutions describing the stability and evolution of solitary
waves in NLS-type systems is a variational approach. This approach is termed modulation theory and is based on using
an averaged Lagrangian and suitable trial functions.  It has been used to study the NLS equation
\cite{Noel95} and optical media, such as nematic liquid crystals \cite{Noel07}.  Related applications and problems
considered for nematic media include dipole motion, boundary induced motion and the development of bores
\cite{dipole,boun,bore}.

In \S 1 modulation theory is developed for the colloidal equations studied here.  A hard-sphere colloidal model is used
with the compressibility described by the Carnahan-Starling formula.  Semi-analytical solutions for both one and
two-dimensional solitary waves are developed.  In \S 2 the modulation equations are solved numerically to obtain
semi-analytical power versus propagation constant curves and neutral stability curves.  These curves illustrate the
multiplicity of the solitary wave solution branches and the regions of parameter space in which unstable solutions
occur.  An excellent comparison between the semi-analytical and numerical solutions is obtained.
In \S 3 conclusions and suggestions for future work are made.

\section{Modulation equations}

Let us consider a coherent light beam (laser light) propagating through
a colloidal suspension of dielectric hard spheres whose diameter is much smaller
than the wavelength of the light and whose refractive index is slightly higher than
 that of the medium in which they are suspended.  With these assumptions,
in non-dimensional form the equations governing the nonlinear propagation of the
beam through the colloidal suspension are (see \cite{optexpr,yuri})
\begin{eqnarray}
 && i \frac{\partial u}{\partial z} + \frac{1}{2} \nabla^{2} u + \left( \eta - \eta_{0}
\right) u  =  0,  \ \ |u|^{2}  =  g(\eta) - g_0, \label{e:elec} \\
 && {\rm with} \ \ g(\eta)  =  \frac{3-\eta}{(1-\eta)^{3}} + \ln \eta, \ \ g_0=g(\eta_0). \nn
\end{eqnarray}
Here $u$ is the envelope of the electric field of the light and $\eta$ is the
packing fraction of the colloid particles, with $\eta_{0}$ the background
fraction.  Losses due to Rayleigh scattering have been neglected as these losses
are small in the limit of the particle diameter being much smaller than the
wavelength of the light \cite{yuri}.  The Carnahan-Starling compressibility
approximation has been used.  Alternative models for the compressibility alter the form of $g$ in (\ref{e:elec}).
The Carnahan-Starling approximation is valid up to the solid-fluid transition, which occurs
at $\eta=\sqrt{2}\pi /9 \approx 0.496$ in a hard-sphere fluid, see \cite{hansen}.

The colloid equations (\ref{e:elec}) have the Lagrangian formulation
\begin{eqnarray}
 && L  =  i \left( u^{*}u_{z} - uu^{*}_{z} \right) - |\nabla u|^{2} + 2\left( \eta
- \eta_{0} \right) |u|^{2} - \frac{4 - 2\eta}{(1-\eta)^{2}}\label{e:lag} \\
&& \mbox{} +\frac{4 - 2\eta_{0}}{(1-\eta_{0})^{2}} - 2\eta \ln \eta + 2\eta_{0} \ln \eta_{0}
+ 2\left( \eta - \eta_{0}\right) ( 1 + g_0). \nn
\end{eqnarray}
Here the asterisk superscript denotes the complex conjugate.
The colloid equations (\ref{e:elec}) possess solitary wave solutions.
However the solitary wave solutions, in both 1-D and 2-d, have only been
found numerically \cite{yuri,poland}, with no analytical solution known.  An approximate
technique which has been found to be useful in these cases for which there is no
known analytical solution on which to base a perturbation analysis is the use
of suitable trial functions in an averaged Lagrangian formulation \cite{Noel95}.  In the
special case in which there is an analytical solitary wave, or periodic wave, solution
this approximate technique is the same as the modulation theory or averaged Lagrangian technique of
Whitham \cite{whitham} and other, related, perturbation techniques \cite{Noel95}.
In this sense this use of trial functions in an averaged Lagrangian is an extension
of Whitham modulation theory.  The approximate trial function method has been applied
to many problems in nonlinear optics and has been found to give solutions in excellent
agreement with numerical and experimental results \cite{Noel07,waveguide}.  The key
is to find a good approximation to the solitary wave solution of a given equation.  However,
in many situations, the evolution of the beam, in particular its position and velocity, are
independent, or nearly so, of the form of the trial function used for the solitary wave profile
\cite{tcollocal,benpoland,ben2}.  This use of trial functions in an averaged Lagrangian,
termed modulation theory in analogy with the modulation theory of Whitham \cite{whitham},
will be used to analyse solitary wave solutions of the present colloid equations.

\subsection{One spatial dimension}

Let us first consider the solitary wave solution of the 1-D form
of the colloid equations (\ref{e:elec}).  The solitary wave solution
of the nonlinear Schr\"odinger (NLS) equation in 1-D has a $\sech$ profile.  In
the small colloid concentration limit $\eta \ll 1$ the colloid equations (\ref{e:elec})
reduce to the NLS equation.  Therefore, let us use the trial functions
\begin{equation}
 u = a \sech \frac{x - \xi}{w} \: e^{i\sigma + iV(x-\xi)} + ig e^{i\sigma + iV(x-\xi)},
\quad \eta = \eta_{0} + \alpha \sech^{2} \frac{x - \xi}{\beta}\label{e:trial1d}
\end{equation}
for the electric field and colloid fraction to analyse the evolution of an initial beam
to a steady solitary wave solution.  The parameters are all functions of $z$.
They are the amplitudes, $a$ and $\alpha$, and the widths, $w$ and $\beta$, of the electric field and colloid
fraction, respectively.  $\sigma$ is the propagation constant of the solitary wave, while
$\xi'=V$ is the velocity of the solitary wave.  Lastly, $g$ is the amplitude of the radiation bed on
which the beam sits.  The first term in the electric field solitary
wave is a varying NLS-type solitary wave.  The second term represents the low wavenumber
(long wavelength) radiation that accumulates under an evolving beam which is not an
exact solitary wave solution.  The origin of this low wavenumber radiation is explained
in detail in \cite{Noel95}.  However, its origin can be deduced from the
group velocity $c_{g} = k$ for linear waves of wavenumber $k$ for the NLS-type equation (\ref{e:elec}).
It can be seen that low wavenumber waves have low group velocity and so accumulate under the beam as it
evolves.  This flat shelf of radiation under the beam matches to shed radiation of
non-zero wavenumber which propagates away from the beam and so allows it to settle to
a steady state \cite{Noel95}.  The flat shelf of radiation cannot extend indefinitely
and so is assumed to have length $\ell$, so that $g$ is non-zero in $\xi - \ell/2 \le x
\le \xi + \ell/2$ \cite{Noel95}.  Finally the flat shelf under the beam is $\pi/2$ out
of phase with it, which accounts for the $i$ multiplying $g$ in the trial function
for $u$ \cite{Noel95}.  As previously mentioned, the flat shelf links with the shed diffractive
radiation which propagates away from the beam.  The effect of this shed radiation
could be included \cite{Noel95}.  However, for the present analysis the effect
of this shed radiation is not needed.

Substituting the trial functions (\ref{e:trial1d}) into the Lagrangian (\ref{e:lag})
and averaging it by integrating in $x$ over the infinite domain results in the averaged
Lagrangian
\begin{eqnarray}
 &&{\cal L}  =  -2\left( 2a^{2}w + \ell g^{2} \right) ( \sigma ' - V\xi' + V^{2}/2)
 - 2\pi awg' + 2\pi gwa'  \label{e:avlag1d} \\
&& +2\pi gaw' - \frac{2}{3} \frac{a^{2}}{w} + 4\alpha a^{2} \Omega_{1}(w,\beta) - \beta \Xi_{1} (\alpha)
- 4 \beta \Theta_{1}(\alpha)
\nn \\
&& + 4\alpha \beta \left( 1 + g_0\right), \ \ {\rm where} \nn \\
&& \Omega_{1}(w,\beta)  =  \int_{0}^{\infty} \sech^{2} \frac{\zeta}{\beta} \sech^{2} \frac{\zeta}{w} \: d\zeta,
\nn \\
&& \Xi_{1}(\alpha)  =  2\int_{0}^{\infty} \left[ \frac{4 - 2\eta_{0} - 2\alpha \sech^{2} \zeta}
{(1 - \eta_{0} - \alpha \sech^{2} \zeta)^{2}} - \frac{4 - 2\eta_{0}}{(1 - \eta_{0})^{2}} \right] \: d\zeta,
\nn \\
&& \Theta_{1}(\alpha)  =   \int_{0}^{\infty} \left[ \eta_{0}\ln ( 1 + \frac{\alpha}{\eta_{0}}
\sech^{2} \zeta ) + \alpha \sech^{2} \zeta \ln \left( \eta_{0} + \alpha \sech^{2} \zeta \right) \right] \: d \zeta. \nn
\end{eqnarray}
Taking variations of this averaged Lagrangian with respect to the solitary wave parameters
gives the modulation equations
\begin{eqnarray}
 && \frac{d}{dz} \left( 2a^{2}w + \ell g^{2} \right)  =  0, \label{e:mass} \\
 && \pi \frac{d}{dz} aw  =  \ell g ( \sigma' - V \xi ' +  V^{2}/2), \label{e:aw} \\
 && \pi \frac{dg}{dz}  =  \frac{2a}{3w^{2}} - \frac{2\alpha a}{w}( \Omega_{1} -
w \Omega_{1w}), \label{e:gp} \\
 && \frac{d\sigma}{dz} - V \frac{d\xi}{dz} + \frac{1}{2}V^{2}  =  - \frac{1}{2w^{2}}
 + \frac{\alpha}{w} ( 2\Omega_{1} - w \Omega_{1w}), \label{e:sigmap} \\
 && \frac{d}{dz} \left( 2a^{2}w + \ell g^{2} \right) V  =  0, \ \
 \frac{d\xi}{dz}  =  V, \label{e:xip} \\
 && 4a^{2}\Omega_{1} - \beta \Xi_{1\alpha} - 4\beta \Theta_{1\alpha} + 4 \beta \left( 1 +
 g_0 \right)  =  0, \label{e:alpha} \\
 && 4\alpha a^{2}  \Omega_{1\beta} - \Xi_{1} - 4\Theta_{1} + 4\alpha \left( 1 +
 g_0 \right)  =  0. \label{e:beta}
\end{eqnarray}
Equation (\ref{e:mass}) is conservation of mass, while the first of equations (\ref{e:xip}) is
conservation of momentum, in the sense of invariances of the Lagrangian (\ref{e:lag}), see \cite{newell}.
However, in the present optical context (\ref{e:mass}) corresponds physically to conservation
of optical power.

A steady solitary wave solution of (\ref{e:elec}) will not shed radiation, so that $g=0$ at the fixed
point of the modulation equations.  Also, at the steady state we can set $V=\xi=0$ for convenience.  Note that
a simple transformation generates non-stationary solitary waves (with $V$ non-zero) with an unchanged profile.
Setting $g=V=\xi=0$ in the modulation equations gives
\begin{eqnarray}
 &&  1-3\alpha w (\Omega_{1}- w\Omega_{1w})=0, \label{e:gps} \\
 &&  \sigma' + \frac{1}{2w^{2}}  - \frac{\alpha}{w} ( 2\Omega_{1} - w \Omega_{1w}) =0, \label{e:sigmaps} \\
 && 4a^{2}\alpha(\Omega_1-\beta\Omega_{1\beta})-\beta(\alpha\Xi_{1\alpha} - \Xi_{1})
 -4\beta(\alpha \Theta_{1\alpha}-\Theta_1)=0, \label{e:alphas} \\
 && 4\alpha a^{2} \Omega_{1 \beta} - \Xi_{1} - 4\Theta_{1} + 4\alpha \left( 1 +
 g_0 \right)  =  0. \label{e:betas}
\end{eqnarray}
These four transcendental equations in the five unknowns $a$, $\alpha$, $w$, $\beta$ and $\sigma'$
represent a two-parameter family of solitary waves which depend on $\eta_{0}$.  The optical power is defined by
\begin{equation}
P=\int_{-\infty}^{\infty} |u(x)|^2 \: dx. \label{power}
\end{equation}
Using the trial function (\ref{e:trial1d}) in (\ref{power}) gives the power of the 1-D semi-analytical solitary wave as
\begin{equation}
P=\int_{-\infty}^{\infty} a^2\sech^2{x\over w} \: dx=2a^2w. \label{power1d}
\end{equation}
Semi-analytical power versus propagation constant curves are described by the solution of (\ref{e:gps})--(\ref{e:betas}).
Stable solution branches occur for $P_{\sigma}>0$ (see \cite{yuri}).  Hence solitary waves of neutral stability have
the property that $P_{\sigma}=0$.  Adding this condition to the equations (\ref{e:gps})--(\ref{e:betas}) gives
a set of five equations for the five unknowns, for given $\eta_{0}$.  Hence, the curves of neutral stability are lines
in the $\sigma$ versus $\eta_0$ plane.  The relevant sets of transcendental equations for both the power versus
propagation constant curves and the lines of neutral stability were solved using a nonlinear equation solver from
the IMSL library.

\subsection{Two spatial dimensions}

The modulation equations of the previous subsection for the evolution of a beam in
a 1-D can be extended to a 2-D beam.  In this case the appropriate trial functions are
\begin{eqnarray}
 && u = a \sech \frac{\phi}{w} \: e^{i\sigma + iU(x-\xi_{x}) + iV(y-\xi_{y})} +
ig e^{i\sigma + iU(x-\xi_{x}) + iV(y - \xi_{y})}, \label{e:trial2d} \\
&& \eta = \eta_{0} + \alpha \sech^{2} \frac{\phi}{\beta},
\ \ {\rm where} \ \ \phi = \sqrt{(x-\xi_{x})^{2} + (y - \xi_{y})^{2}}. \nn
\end{eqnarray}
Substitution of these trial functions into the Lagrangian (\ref{e:lag}) and integrating
in $x$ and $y$ from $-\infty$ to $\infty$ results in
the averaged Lagrangian
\begin{eqnarray}
 && {\cal L}  =  -2\left( I_{2}a^{2}w^{2} + \Lambda g^{2} \right) ( \sigma ' - U\xi_{x}'
 - V\xi_{y}' + U^{2}/2 + V^{2}/2)
 \label{e:avlag2d} \\
&& -2I_{1}aw^{2}g'+ 2I_{1}gw^{2}a'
+ 4I_{1}awgw' - I_{22}a^{2} + 2\alpha a^{2}\Omega_{2}(w,\beta) \nn \\
&&  - \beta^{2} \Xi_{2}(\alpha)- 2\beta^{2}\Theta_{2}(\alpha)
+ 2I_{2}\alpha \beta^{2} \left( 1 + g_{0} \right).
\nn
\end{eqnarray}
In this 2-D case the shelf of low wavenumber radiation under the beam forms a
circle of radius $\ell$, so that $g$ is non-zero in the circle
\begin{equation}
0 \le \sqrt{(x-\xi_{x})^{2} + (y -\xi_{y})^{2}} \le \ell.
\label{e:l2d}
\end{equation}
Here $\Lambda = \ell^{2}/2$.  The various integrals involved in this averaged Lagrangian are
\begin{eqnarray}
 && I_{1}  =  \int_{0}^{\infty} \zeta \sech \zeta \: d\zeta = 2C,  \ \
 I_{2} =  \int_{0}^{\infty} \zeta \sech^{2} \zeta \: d\zeta = \ln 2, \nn \\
 && I_{22}  =  \int_{0}^{\infty} \zeta \sech^{2} \zeta \tanh^{2} \zeta \: d\zeta =
\frac{1}{3} \ln 2 + \frac{1}{6}, \nn \\
 && \Omega_{2}(w,\beta)  =  \int_{0}^{\infty} \zeta \sech^{2} \frac{\zeta}{\beta} \sech^{2}
\frac{\zeta}{w} \: d\zeta, \label{e:ints2}  \\
&& \Xi_{2}(\alpha)  =  \int_{0}^{\infty} \zeta \left[ \frac{4 - 2\eta_{0} - 2\alpha \sech^{2} \zeta}
{(1 - \eta_{0} - \alpha \sech^{2} \zeta)^{2}} - \frac{4 - 2\eta_{0}}{(1 - \eta_{0})^{2}} \right] \: d\zeta,
\nn \\
&& \Theta_{2}(\alpha)  =   \int_{0}^{\infty} \zeta \left[ \eta_{0}\ln ( 1 + \frac{\alpha}{\eta_{0}}
\sech^{2} \zeta ) \right. \nonumber \\
& & \left. \mbox{} + \alpha \sech^{2} \zeta \ln \left( \eta_{0} + \alpha \sech^{2} \zeta \right) \right]
\: d\zeta, \nonumber
\end{eqnarray}
where $C$ is the Catalan constant $C=0.915965594\ldots$ \cite{abram}.
The modulation (variational) equations for the averaged Lagrangian (\ref{e:avlag2d}) are
\begin{eqnarray}
&&  \frac{d}{dz} \left( I_{2}a^{2}w^{2} + \Lambda g^{2} \right)  =  0, \label{e:mass2d} \\
 && I_{1} \frac{d}{dz} aw^{2}  =  \Lambda g ( \sigma' - U\xi_{x}' - V\xi_{y}' + U^{2}/2
+ V^{2}/2 ), \label{e:aw22d} \\
&& 2I_{1} \frac{dg}{dz}  =  \frac{I_{22}a}{w^{2}} - \frac{\alpha a}{w^{2}} \left( 2\Omega_{2} -
 w\Omega_{2w} \right), \label{e:gp2d} \\
&& I_{2} \left( \frac{d\sigma}{dz} - U\frac{d\xi_{x}}{dz} - V\frac{d\xi_{y}}{dz} + \frac{1}{2}
U^{2} + \frac{1}{2}V^{2} \right)  =  \nonumber \\
&&  \mbox{} - \frac{I_{22}}{w^{2}} + \frac{\alpha}{2w^{2}} (
4\Omega_{2} - w \Omega_{2w}), \label{e:sigma2d} \\
 && \frac{d}{dz} \left( I_{2}a^{2}w^{2} + \Lambda g^{2} \right) U  =  0, \ \
  \frac{d}{dz} \left( I_{2}a^{2}w^{2} + \Lambda g^{2} \right) V  =  0, \label{e:ymom2d} \\
 && \frac{d\xi_{x}}{dz}  = U, \ \  \frac{d\xi_{y}}{dz}  =  V, \label{e:xiy2d} \\
 && 2a^{2}\Omega_{2} - \beta^{2}  \Xi_{2\alpha} - 2\beta^{2}
 \Theta_{2\alpha} + 2I_{2} \beta^{2} \left( 1 + g_{0} \right)  =  0, \label{e:alpha2d} \\
 && \alpha a^{2}  \Omega_{2\beta} - \beta \Xi_{2} - 2\beta \Theta_{2}
 + 2I_{2}\alpha \beta \left( 1 + g_{0} \right)  =  0. \label{e:beta2d}
\end{eqnarray}
At the steady-state we have $g=V=U=\xi=0$ and the modulation equations become
\begin{eqnarray}
&& I_{22} - \alpha  \left( 2\Omega_{2} -
 w \Omega_{2w} \right)=0, \label{e:sgp2d} \\
 && I_{2}\sigma' + \frac{I_{22}}{w^{2}} - \frac{\alpha}{2w^{2}} (
4\Omega_{2} - w \Omega_{2w}) = 0, \label{e:ssigma2d} \\
&& 2a^{2}\Omega_{2} - \beta^{2} \Xi_{2\alpha} - 2\beta^{2}
 \Theta_{2\alpha} + 2I_{2} \beta^{2} \left( 1 + g_{0} \right)  =  0, \label{e:salpha2d} \\
 && \alpha a^{2} \Omega_{2\beta} - \beta \Xi_{2} - 2\beta \Theta_{2}
 + 2I_{2}\alpha \beta \left( 1 + g_{0} \right)  =  0. \label{e:sbeta2d}
\end{eqnarray}
As in the 1-D case, equations (\ref{e:sgp2d})--(\ref{e:sbeta2d}) give a semi-analytical description
of a two parameter family of 2-D colloidal solitary waves.  The optical power of these solitary
waves is given by
\begin{equation}
P=\int_{0}^{\infty} r|u(r)|^2 \: dr.
\label{power1}
\end{equation}
The power of a semi-analytical 2-D solitary wave is then found by 
substituting the trial function (\ref{e:trial2d}) into (\ref{power1}), giving
\begin{equation}
P=\int_{0}^{\infty} r a^2\sech^2{r\over w} \: dr= a^2w^2 \ln 2.
\end{equation}

\section{Results and discussion}

\begin{figure}[t]
\centering 
\includegraphics[scale=0.9]{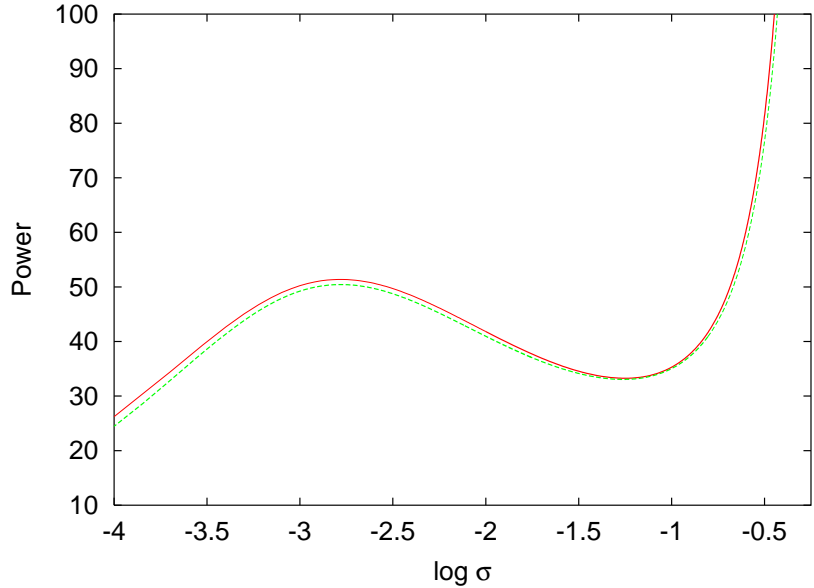}
\caption{The power versus propagation constant ($P$ versus log $\sigma$) curve for the 1-D colloidal solitary wave.
Shown are the semi-analytical (solid lines, red) and numerical (dashed lines, green) solutions.
The background fraction is $\eta_0=1\times 10^{-3}$. \label{fig1}}
\end{figure}

\begin{figure}[t]
\centering 
\includegraphics[scale=0.9]{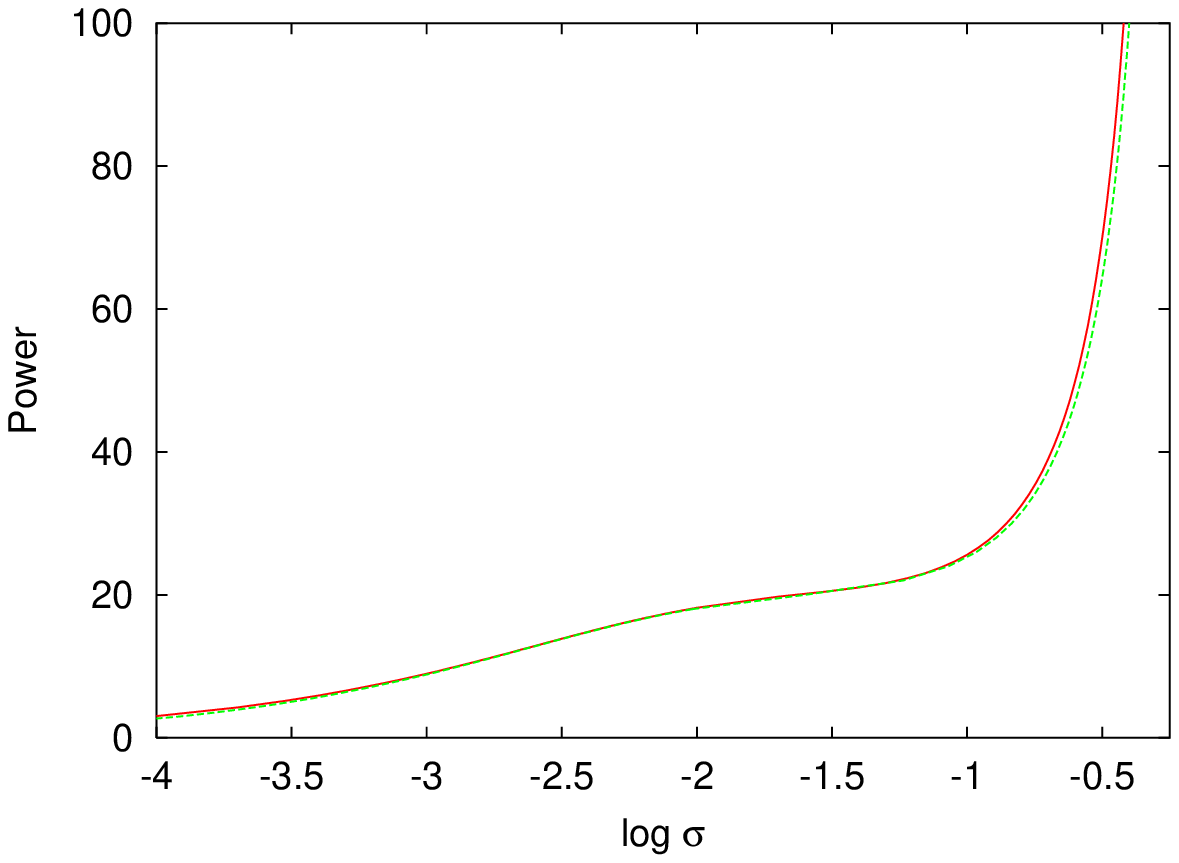}
\caption{The power versus propagation constant ($P$ versus log $\sigma$) curve for the 1-D colloidal solitary wave.
Shown are the semi-analytical (solid lines, red) and numerical (dashed lines, green) solutions.
The background fraction is $\eta_0=1\times 10^{-2}$. \label{fig2}}
\end{figure}

In this section the semi-analytical solutions for colloidal solitary waves are compared with numerical
solutions.  Semi-analytical estimates for the power versus propagation constant and neutral stability curves
are both found.  The numerical solutions in 1-D are obtained by analytically integrating the steady-state version
of the governing equation to obtain an energy conservation law.  The energy conservation law can then be
numerically integrated to obtain exact solitary wave profiles on both the stable and unstable solution branches
or the power versus propagation constant curves.  In 2-D the numerical solutions were obtained using the imaginary time
iterative method suitable for obtaining numerically exact solitary wave profiles \cite{yang08}.  Note that the imaginary time
method cannot be used to obtain numerical solutions on an unstable solution branch.

\subsection{One spatial dimension}

Figure \ref{fig1} shows the power versus propagation constant ($P$ versus ${\rm log} \sigma$) curve for the
1-D colloidal solitary wave.  The background packing fraction is $\eta_0=1\times 10^{-3}$.  Shown are the
semi-analytical solution (\ref{e:gps})--(\ref{e:betas}) and the numerical solution.  The figure uses the same
parameters as Figure 3 of \cite{yuri}.  It can be seen that there is an excellent comparison between the
semi-analytical and numerical solutions.  The figure shows two stable branches, separated by a middle, unstable
solution branch.  The solitary waves on the low power stable branch have low amplitudes and large widths, whilst
those on the high power stable branch have large amplitudes and smaller widths.  The semi-analytical solution
indicates that the middle unstable branch exists for
\begin{eqnarray}
&& -2.77<{\rm log} \sigma <-1.25 \ \ {\rm and}  \ \  33.26<P<51.37 \label{branch} \\
&& 1.31<a<2.29 \ \ {\rm  and}  \ \ 4.29\times 10^{-3} < \alpha < 0.19. \nn
\end{eqnarray}

Figure \ref{fig2} shows the power versus propagation constant ($P$ versus $\log \sigma$) curve for the
1-D colloidal solitary wave for the background packing fraction $\eta_0=1\times 10^{-2}$.  Shown are the
semi-analytical solution (\ref{e:gps})--(\ref{e:betas}) and the numerical solution.  In this example the background
packing fraction has been increased, which eliminates the bistability.  The propagation constant versus power
curve now has a single, stable solution branch.  Again the comparison between the semi-analytical and numerical
solutions is excellent.  Figures \ref{fig1} and \ref{fig2} illustrate that the semi-analytical solution is
extremely accurate and hence is highly suitable for obtaining accurate results relating to the stability and
other properties of 1-D colloidal solitary waves.

\begin{figure}[t]
\centering
\includegraphics[scale=0.9]{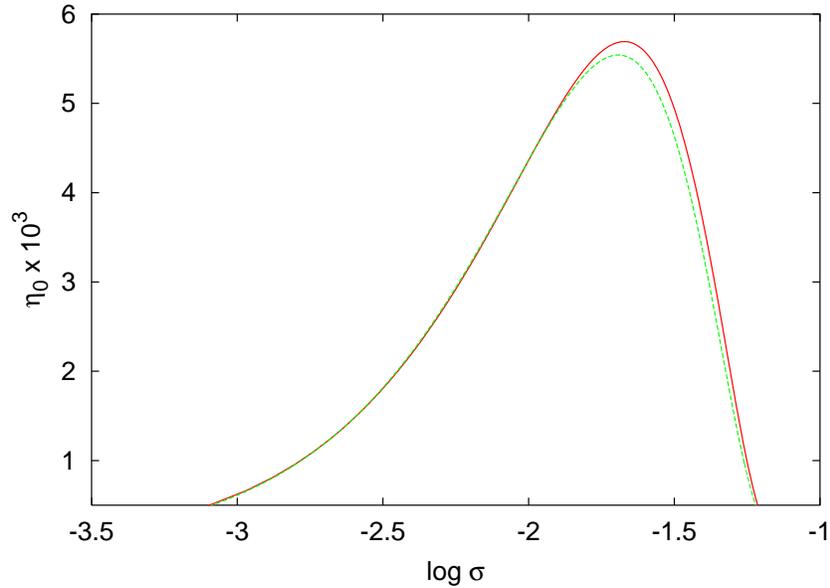}
\caption{\label{fig3}The neutral stability curve in the propagation constant-background packing fraction
plane ($\log \sigma$,$\eta_{0}$) for the 1-D colloidal solitary wave. Shown are the
semi-analytical (solid lines, red) and numerical (dashed lines, green) solutions.}
\end{figure}

Figure \ref{fig3} shows the neutral stability curve in the propagation constant-background packing fraction
plane ($\log \sigma$, $\eta_0$) for the 1-D colloidal solitary wave.  Shown are the semi-analytical and numerical
solutions.  The semi-analytical and numerical neutral stability curves were found by solving the condition
$P_{\sigma}=0$ using the expressions for both the semi-analytical and numerical power.
The region under the curves represents parameter values corresponding to the middle, unstable branch of solitary
wave solutions.  The figure shows that as the background packing fraction increases,
the region of parameter space in which unstable solutions occur is reduced and then eliminated.
The parameters of the solitary wave with neutral stability at the turning point are
\begin{eqnarray}
&& ({\rm log} \sigma, \eta_0, a,\alpha)=(-1.67,5.69 \times 10^{-3},1.71,5.65\times 10^{-2}),
\label{degen} \\
&& ({\rm log} \sigma, \eta_0, a,\alpha)=(-1.69,5.54 \times 10^{-3},1.72,5.79\times 10^{-2}),
\label{degennum}
\end{eqnarray}
where (\ref{degen}) and (\ref{degennum}) are the semi-analytical and numerical solitary waves, respectively.
It can be seen that the semi-analytical prediction of the background packing fraction at the turning point
is extremely accurate, with less than $3\%$ error.  Hence bistable behaviour only occurs in
1-D geometry for $\eta_0\le 5.54 \times 10^{-3}$ and a single stable solution branch exists for background
packing fractions greater than this value.  This is consistent with the bistable curve of Figure \ref{fig1} and
the monotone stability seen in Figure \ref{fig2}.

\subsection{Two spatial dimensions}

\begin{figure}[t]
\centering 
\includegraphics[scale=0.9]{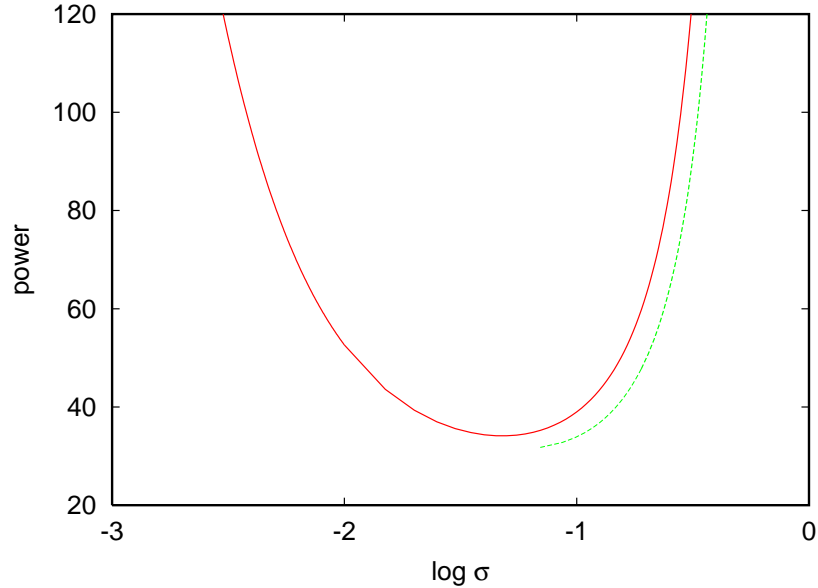}
\caption{The power versus propagation constant ($P$ versus $\log \sigma$) curve for the 2-D colloidal solitary wave.
Shown are the semi-analytical (solid lines, red) and numerical (dashed lines, green) solutions.
The background fraction is $\eta_0=1\times 10^{-3}$. \label{fig4a}}
\end{figure}

Figure \ref{fig4a} shows the power versus propagation constant ($P$ versus $\log \sigma$) curve for the 2-D
colloidal solitary wave.  The background packing fraction is $\eta_0=1\times 10^{-3}$.  Shown are the
semi-analytical solution (\ref{e:gps})--(\ref{e:betas}) and the numerical solution.  This figure is qualitatively
different from the 1-D case of Figure \ref{fig1} as the bistable behaviour is absent.  There are two solution
branches, one stable and the other unstable.  The stable solution branch corresponds to solitary waves of large
amplitude.  The semi-analytical theory predicts that the stable branch occurs for
\begin{equation}
\sigma >0.06, \ \  a>3.81 \ \ {\rm and} \ \  \alpha>0.46.
\end{equation}
The comparison between the semi-analytical and numerical solutions is very good; there is a $17\%$ difference in the
power and a $6\%$ difference in the amplitude $a$ at $\sigma=0.18$.  Note that the maximum packing fraction of the
stable solitary wave of minimum amplitude is $\eta=\alpha+\eta_0=0.461$, which is close to the value
$\eta\approx 0.496$ at which solidification occurs in the hard sphere model.  Hence this branch of stable
solitary waves is unlikely to be realised in real colloidal suspensions.

\begin{figure}[t]
\centering 
\includegraphics[scale=0.9]{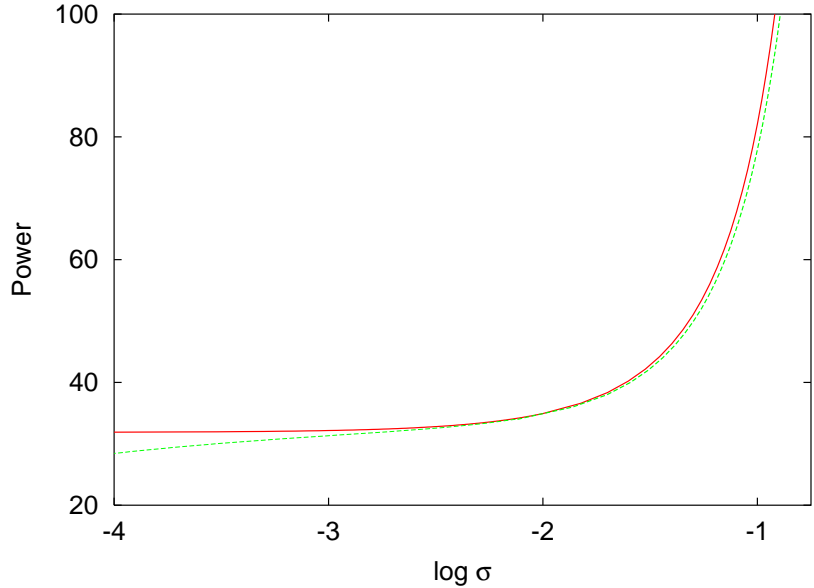}
\caption{The power versus propagation constant ($P$ versus log $\sigma$) curve for the 2-D colloidal solitary wave.
Shown are the semi-analytical (solid lines, red) and numerical (dashed lines, green) solutions.
The background fraction is $\eta_0=0.3$. \label{fig6}}
\end{figure}

Figure \ref{fig6} shows the power versus propagation constant ($P$ versus $\log \sigma$) curve for the 2-D
colloidal solitary wave for the background packing fraction $\eta_0=0.3$.  Shown are the semi-analytical
solution (\ref{e:sgp2d})--(\ref{e:sbeta2d}) and the numerical solution.  For this much larger value of the
background packing fraction there is only one stable solution branch.  Hence, multiple solution branches occur
in both the 1-D and 2-D cases only if the background packing fraction is small enough.  The comparison between
the semi-analytical and numerical solutions is again excellent, except for some slight variation for $\log \sigma <-3$.
For larger powers the amplitudes of the electric field and colloidal fraction pulses increase; beyond $P\approx 50$ the
hard-sphere colloid model predicts solidification.  As in the 1-D case the comparison between the semi-analytical
and numerical solutions is excellent and confirms the suitability of semi-analytical methods for understanding
colloidal wave stability.

\begin{figure}[t]
\centering 
\includegraphics[scale=0.9]{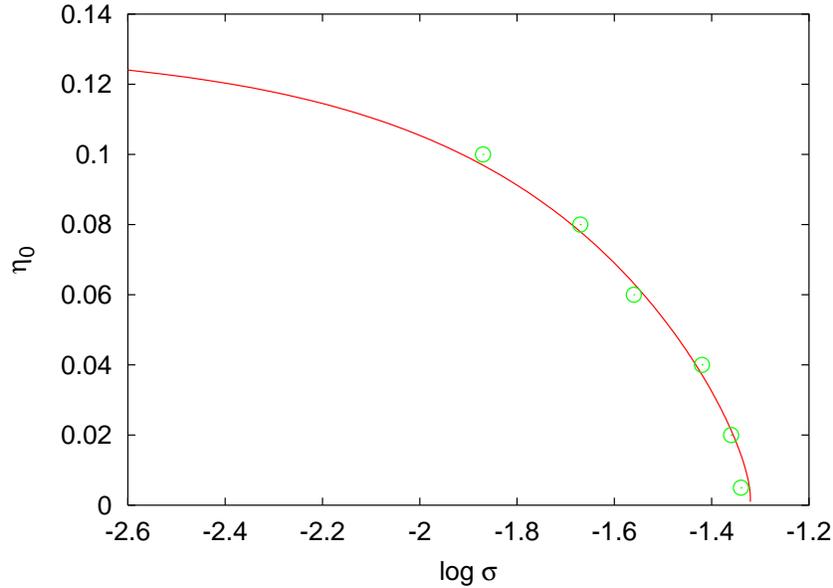}
\caption{\label{fig5}The neutral stability curve in the propagation constant-background packing fraction
plane ($\log \sigma$, $\eta_0$) for the 2-D colloidal solitary wave.  Shown is the semi-analytical solution
(solid line, red) and numerical solutions (circles, green).}
\end{figure}

Figure \ref{fig5} shows the neutral stability curve in the propagation constant-background packing fraction
plane ($\log \sigma$, $\eta_0$) for the 2-D colloidal solitary wave.  Shown are the semi-analytical and numerical solutions.
The semi-analytical solution is found  by solving
(\ref{e:sgp2d})--(\ref{e:sbeta2d}) and the condition $P_{\sigma}=0$. The numerical solutions are found using the
imaginary time iterative method, which is used to find stable solitary wave solutions. For a given
background packing fraction, the numerically obtained solution of minimum power is shown.
This solution lies close to the turning point on the propagation constant versus power curve, and is
a numerical estimate of the neutrally stable solitary wave.
 The region to the left of the semi-analytical curve
represents parameter values corresponding to the unstable branch of solitary wave solutions.  The curve indicates
that multiple solitary wave solution branches only occur for $\eta_0 <0.125$ and that for larger background packing
fractions a single stable solution branch occurs.  As the semi-analytical curve of neutral stability is traversed from right
to left the amplitude of the colloid beam decreases.
It can be seen that there is an excellent comparison between the semi-analytical curve and numerical estimates for the neutrally
stable solitary wave. No numerical estimates near $\eta_0=0.125$ are presented due to computational
difficulties in resolving marginally unstable
and stable cases. This is related to the fact that, in this limit, the propagation
constant versus power curve becomes very nearly flat.

\begin{figure}[t]
\centering 
\includegraphics[scale=0.9]{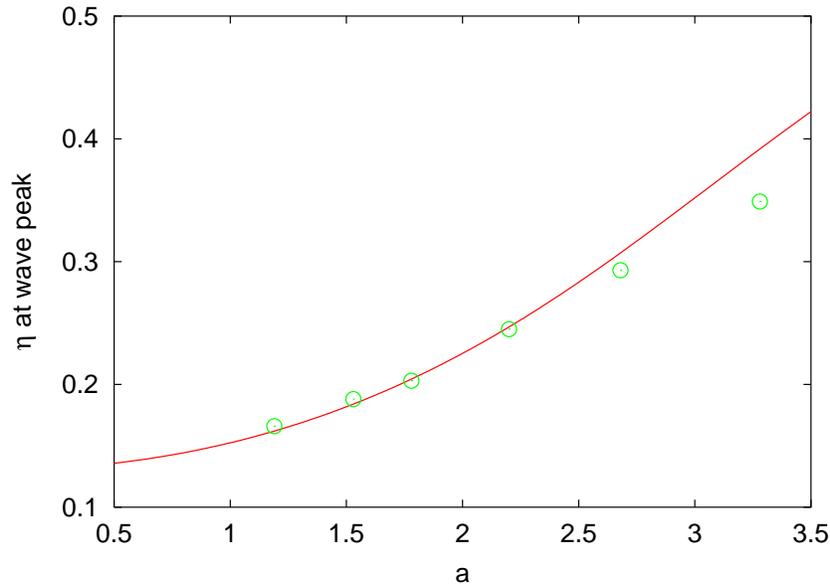}
\caption{\label{fig7}The neutral stability curve in the amplitude-maximum packing fraction
plane ($a$, $\eta$) for the 2-D colloidal solitary wave. Shown is the semi-analytical solution
(solid line, red) and numerical solutions (circles, green).}
\end{figure}

Figure \ref{fig7} shows the neutral stability curve in the amplitude-maximum packing fraction
plane ($a$, $\eta$) for the 2-D colloidal solitary wave.  Shown are the semi-analytical and numerical solutions.
The semi-analytical solution is found by solving
(\ref{e:sgp2d})--(\ref{e:sbeta2d}) and the condition $P_{\sigma}=0$.  The numerical solutions are found using the
imaginary time iterative method, which is used to find stable solitary wave solutions.
The region below the semi-analytical curve
represents parameter values corresponding to the unstable branch of solitary wave solutions.  This alternative
view of the neutral stability curve shows the peak electric field intensity and colloid packing fraction of the
neutrally stable solitary waves.  As the curve is traversed from left to right the background packing fraction
decreases from $\eta_0=0.125$ to zero. Hence stable solitary waves have large maximum packing fractions
when $\eta_0$ is small. As solidification occurs for $\eta \ge 0.496$, it can be
seen that physically realistic, but stable, solitary waves occur for a small range of
amplitudes in the small background packing fraction limit.  As for figure {\ref{fig5}, the 
comparison between semi-analytical and numerical solutions is excellent.

\section{Conclusions}

In this paper a semi-analytical model for colloidal solitary waves described by the hard-sphere gas model with the
Carnahan-Starling approximation has been developed.  Power versus propagation constant curves have been derived for
both 1-D and 2-D geometries, with an excellent comparison with numerical solutions obtained.  The neutral stability
curves show that multiple solution branches only occur for low background packing fractions.  The critical
background packing fractions at which multiplicity is lost are well determined by the semi-analytical theory.

Future work using the semi-analytical model will involve a number of directions.  Firstly, the semi-analytical theory
will be used to examine the unsteady evolution of colloidal beams and the development of colloidal undular bores.
This approach has been been shown to be successful in tackling these problems in related optical media, such as
nematic liquid crystals \cite{bore}.  Secondly, the semi-analytical tools developed here can be used to analyse
other possible colloidal models which use different gas compressibility laws or are composed of multiple
nanoparticle species with different optical properties \cite{matus}.  This will help our understanding of the stability
regimes for 2-D beams in experimental scenarios involving physically important colloidal materials.

\section{Acknowledgements}

This research was supported by the Royal Society of London under grant JP090179.

\noindent Received August 2010; revised February 2011; revised September 2011.\\[2mm]
http://monotone.uwaterloo.ca/$\sim$journal/


\footnotesize
\begin{thebibliography}{99}

\bibitem{abram}  M. Abramowitz and I.A. Stegun, {\em Handbook of Mathematical
  Functions with Formulas, Graphs and Mathematical Tables,} Dover
     Publications, Inc., New York (1972).

\bibitem{boun}  A. Alberucci, G. Assanto, D. Buccoliero, A. Desyatnikov,
T.R. Marchant and N.F. Smyth, ``Modulation analysis of boundary
induced motion of nematicons,'' {\em Phys.\ Rev.\ A,} {\bf 79}, 043816 (2009).

\bibitem{ashkin} A. Ashkin, J.M. Dziedzic, J.E. Bjorkholm and S. Chu, ``Observation of a single-beam gradient
force optical trap for dielectric particle,'' {\em Opt.\ Lett.,} {\bf 11}, 288 (1986).

\bibitem{bore}  G. Assanto, T.R. Marchant and N.F. Smyth, ``Collisionless shock resolution
     in nematic liquid crystals,'' {\em Phys.\ Rev.\ A,} {\bf 78}, 063808 (2008).

\bibitem{benpoland}  G. Assanto, B.D. Skuse and N.F. Smyth, ``Optical path
control of spatial optical solitary waves in dye-doped nematic liquid
crystals,'' {\em Photon.\ Lett.\ Poland,} {\bf 1}, 154--156 (2009).

\bibitem{waveguide}  G. Assanto, A.A. Minzoni, M. Peccianti and N.F. Smyth,
``Optical solitary waves escaping a wide trapping potential in nematic liquid
crystals: modulation theory,'' {\em Phys.\ Rev.\ A,} {\bf 79}, 033837 (2009).

\bibitem{ben2} G. Assanto, B.D. Skuse and N.F. Smyth, ``Solitary wave propagation and steering
through light-induced refractive potentials,'' {\em Phys.\ Rev.\ A,} {\bf 81}, 063811 (2010).

\bibitem{daoud}  M. Daoud and C.E. Williams eds., {\em Soft Matter Physics}, Springer (1999).

\bibitem{elgan} R. El-Ganainy, D.N. Christodoulies, C. Rotschild and M. Segev, ``Soliton dynamics and
self-induced transparency in nonlinear suspensions,'' {\em Opt.\ Express,} {\bf 15}, 10207--10218 (2007).

\bibitem{elgan09} R. El-Ganainy, D.N. Christodoulies, E.M. Wright, W.M. Lee and K. Dholakia, ``Nonlinear
optical dynamics in nonideal gases of interacting colloidal nanoparticles,'' {\em Phy.\ Rev.\ A,} {\bf 80}, 053805 (2009).

\bibitem{dipole}  C. Garc\'{\i}a-Reimbert, A.A. Minzoni, T.R. Marchant, N.F. Smyth
     and A.L. Worthy, ``Dipole soliton formation in a nematic liquid
     crystal in the nonlocal limit,'' {\em Physica D,} {\bf 237}, 1088--1102 (2008).

\bibitem{grier}  D.V. Grier, ``A revolution in optical manipulation,'' {\em Nature,} {\bf 424}, 810 (2003).

\bibitem{hansen}  J.P. Hansen and I.R. McDonald, {\em Theory of Simple Liquids},
  Academic Press, London (1976).

\bibitem{Noel95}   W.L. Kath and N.F. Smyth, ``Soliton evolution and radiation
     loss for the nonlinear Schr\"odinger equation,'' {\em Phys.\ Rev.\ E,} \textbf{51}, 1484--1492 (1995).

\bibitem{newell}  D.J. Kaup and A.C. Newell, ``Solitons as particles,
   oscillators, and in slowly changing media:  a singular perturbation
   theory,'' {\em Proc.\ Roy.\ Soc.\ Lond.\ A,} {\bf 361}, 413--446 (1978).

\bibitem{lee} W.M. Lee, R. El-Ganainy, D.N. Christodoulies, K. Dholakia and E.M. Wright, ``Nonlinear
optical response of colloidal suspensions,'' {\em Opt.\ Express,} {\bf 17}, 10277--10289 (2009).

\bibitem{optexpr}  M. Matuszewski, W. Krolikowski and Y.S. Kivshar, ``Spatial solitons
and light-induced instabilities in colloidal media,'' {\em Opt.\ Express,}
{\bf 16}, 1371--1376 (2008).

\bibitem{yuri}  M. Matuszewski, W. Krolikowski and Y.S. Kivshar, ``Soliton
interactions and transformations in colloidal media,'' {\em Phys.\ Rev.\
A,} {\bf 79}, 023814 (2009).

\bibitem{poland}  M. Matuszewski, W. Krolikowski and Y.S. Kivshar, ``Bistable solitons in
colloidal media,'' {\em Photon.\ Lett.\ Poland,} {\bf 1}, 4--6 (2009).

\bibitem{matus} M. Matuszewski, ``Engineering optical soliton bistability in colloidal media,''
{\em Phy.\ Rev.\ A,} {\bf 81}, 013820, (2010).

\bibitem{Noel07}  A.A. Minzoni, N.F. Smyth and A.L. Worthy, ``Modulation
     solutions for nematicon propagation in non-local liquid crystals,''
{\em J.\ Opt.\ Soc.\ Amer.\ B,} \textbf{24}, 1549--1556 (2007).

\bibitem{tcollocal}  B.D. Skuse and N.F. Smyth, ``Two-colour vector soliton
    interactions in nematic liquid crystals in the local response regime,''
   {\em Phys.\ Rev.\ A,} {\bf 77}, 013817 (2008).

%\bibitem{vortex}  A.A. Minzoni, N.F. Smyth, A.L. Worthy and Y.S. Kivshar,
%     ``Stabilization of vortex solitons in nonlocal nonlinear media,''
%     {\em Phys.\ Rev.\ A,} {\bf 76}, 063803 (2007).


%\bibitem{angmom}  G. Assanto, N.F. Smyth and A.L. Worthy, ``Two colour, nonlocal
%      vector solitary waves with angular momentum in nematic liquid crystals,''
 %     {\em Phys.\ Rev.\ A,} {\bf 78}, 013832 (2008).

%\bibitem{trap}  G. Assanto, A.A. Minzoni, M. Peccianti and N.F. Smyth,
%``Optical solitary waves escaping a wide trapping potential in nematic liquid
%crystals: modulation theory,'' {\em Phys.\ Rev.\ A,} {\bf 79}, 033837 (2009).

\bibitem{whitham}  G.B. Whitham, {\em Linear and Nonlinear Waves,} J. Wiley and Sons,
New York (1974).

\bibitem{yang08} J. Yang and T.I. Lakoba, ``Accelerated imaginary-time evolution methods
for the computation of solitary waves'',
  {\em Stud. Appl. Math.,} {\bf 120}, 265-292 (2008).

%\bibitem{vecvortex}  Z. Xu, N.F. Smyth, A.A. Minzoni and Y.S. Kivshar,
%``Vector vortex solitons in nematic liquid crystals,'' {\em Opt.\
%Lett.\ } {\bf 34}, pp.\ 1414--1416 (2009).

%\bibitem{twocolnon}  B.D. Skuse and N.F. Smyth, ``Interaction of two colour
%     solitary waves in a liquid crystal in the nonlocal regime,''
%     {\em Phys.\ Rev.\ A,} {\bf 79}, 063806 (2009).

%\bibitem{vecvortex2}  A.A. Minzoni, N.F. Smyth, Z. Xu and Y.S. Kivshar,
%``Stabilization of vortex-soliton beams in nematic liquid crystals,''
%     {\em Phys.\ Rev.\ A,} {\bf 79}, 063808 (2009).

%\bibitem{survey}  G. Assanto, A.A. Minzoni and N.F. Smyth, N.F.,
%``Light self-localization in nematic liquid crystals: modelling solitons in
%nonlocal reorientational media,'' {\em J. Nonlin.\ Opt.\ Phys.\ and Mater.\ }
%{\bf 18}, 657--691 (2009).

%\bibitem{circvortex}  A.A. Minzoni, N.F. Smyth and Z. Xu,
%``Stability of an optical vortex in a circular nematic cell,'' {\em Phys.\
%Rev. A,} {\bf 81}, 033816 (2010).

%\bibitem{KiAg03}
%Y.S. Kivshar and G.P. Agrawal, {\em Optical Solitons. From Fibers to Photonic Crystals},
%Academic Press, San Diego (2003).

%\bibitem{fleischer}  W. Wan, D.V. Dylov, C. Barsi and J.W. Fleischer, ``Dispersive
%shock waves with negative pressure,'' {\em Conference on Lasers and Electro-Optics/International
%Quantum Electronics Conference, OSA Technical Digest (CD),} (Optical Society of America),
%paper IThM5 (2009).

%\bibitem{GuPi74}
%A.V. Gurevich and L.P. Pitaevskii,
%``Nonstationary structure of a collisionless shock wave,''
%{\em Sov.\ Phys., J. Exp.\ Theor.\ Phys.\ 33}, 291--297 (1974).

%\bibitem{el}  G.A. El, V.V. Geogjaev, A.V. Gurevich and A.L. Krylov, ``Decay of an initial
%discontinuity in the defocusing NLS hydrodynamics,'' {\em Physica D,} {\bf 87},
%186--192 (1995).

%\bibitem{forest}  H. Flaschka, M.G. Forest and D.W. McLaughlin, ``Multiphase averaging and
%the inverse spectral solution of the Korteweg-de Vries equation ,'' {\em Comm.\ Pure Appl.\ Math.,}
%{\bf 33}, 739--784 (1980).

%\bibitem{el1}  G.A. El, V.V. Khodorovskii and A.V. Tyurina, ``Determination of boundaries
%of unsteady oscillatory zone in asymptotic expanions of nonintegrable dispersive wave
%equations.'' {\em Phys.\ Lett.\ A,} {\bf 318}, 526--536 (2003).

%\bibitem{el2}  G.A. El, V.V. Khodorovskii and A.V. Tyurina, ``Undular bore transition
%in bidirectional conservative wave dynamics,'' {\em Physics D,} {\bf 206}, 232--251 (2005).

%\bibitem{el3}  G.A. El, ``Resolution of a shock in hyperbolic systems modified by weak
%dispersion,'' {\em Chaos,} {\bf 15}, 037103 (2005).

%\bibitem{fleischer2}  W. Wan, S. Jia and J.W. Fleischer, ``Dispersive superfluid-like shock
%waves in nonlinear optics,'' {\em Nature Physics,} {\bf 3}, 46--51 (2007).

%\bibitem{fleischer3}  C. Barsi, W. Wan, C. Sun and J.W. Fleischer, ``Dispersive shock waves
%with nonlocal nonlinearity,'' {\em Opt.\ Lett.,} {\bf 32}, 2930--2932 (2007).

%\bibitem{trillo}  N. Ghofraniha, C. Conti, G. Ruocco and S. Trillo, ``Shocks in nonlocal media,''
%{\em Phys.\ Rev.\ Lett.,} {\bf 99}, 043903 (2007).

%\bibitem{conti}  C. Conti, A. Fratalocchi, M. Peccianti, G. Ruocco and S. Trillo, ``Observation
%of a gradient catastrophe generating solitons,'' {\em Phys.\ Rev.\ Lett.,} {\bf 102},
%083902 (2009).

%\bibitem{GrSm86}
%R.H.J. Grimshaw and N.F. Smyth,
%``Resonant flow of a stratified fluid over topography,''
%{\em J. Fluid Mech.\ 169}, 429--464 (1986).

\end{thebibliography}
\end{document}